\documentclass[11pt,a4paper]{article}
\usepackage{amsmath,amssymb}
\usepackage{epsfig,graphicx}
\usepackage{subfigure}
\usepackage{graphicx}
\usepackage{rotating}
\usepackage{bm}
\usepackage{axodraw}
\usepackage{color}
\usepackage[sort&compress,square]{natbib}

\begin{document}
\date{\mbox{ }}
\title{{\normalsize DESY 06-190; IPPP/06/69; DCPT/06/138\hfill\mbox{}\\
October 2006\hfill\mbox{}}\\
\vspace{1.5cm} \textbf{Entropy Growth and the Dark Energy\\
Equation of State}\footnote{
Based on a talk given by J. Jaeckel at the QUARKS 2006 conference.}\\
[5mm]}
\author{W. Buchm\"uller$^a$\footnote{{\bf e-mail}: 
wilfried.buchmueller@desy.de},
J. Jaeckel$^{a,b}$\footnote{{\bf e-mail}:
jjaeckel@mail.desy.de}
\\
$^a$ \small{\em Deutsches Elektronen Synchrotron} \\
\small{\em Notkestrasse 85, 22607 Hamburg, Germany} \\
$^b$ \small{\em Center for Particle Theory, Durham University} \\
\small{\em Durham DH1 3LE, United Kingdom} }
\date{}
\maketitle

\begin{abstract}
\noindent
We revisit the conjecture of a generalized second law of thermodynamics which 
states that the combined entropy of matter and horizons must grow.
In an expanding universe a generalized second law restricts the equation of 
state. In particular,
it conflicts with long phases of a phantom, $w < -1$, equation of state.
\end{abstract}

\section{Introduction}
Observations indicate that our universe is in a phase of
accelerated expansion
\cite{Riess:2004nr,Spergel:2003cb,Tegmark:2003ud}.  Some
mysterious dark energy appears to drive this  acceleration. There
exist various attempts to explain this phenomenon, most notably
the (in)famous cosmological constant \cite{zeld,Weinberg:1989cp}
and a scalar field dubbed ``quintessence''
\cite{Wetterich:fm,Ratra:1987rm,Caldwell:1997ii}. Both contribute
a ``dark energy'' component to the cosmic energy density which can be described
by a perfect fluid with an equation of state
\mbox{$w=\frac{p}{\rho}=\frac{T-V}{T+V}$}. 

Accelerated expansion requires
$w<-\frac{1}{3}$. It is readily seen that for a
positive potential $V$ and positive kinetic energy $T$, one has $-1\leq
w\leq1$. The cosmological constant, having no kinetic energy, has
$w=-1$.
Recently, however, also the case $w< -1$ has attracted a
lot of attention, because it appears  slightly
favored by supernovae data \cite{Caldwell:2003vq}.
A typical realization of  $w<-1$ is provided by a scalar field with an
inverted sign for the kinetic term \cite{Caldwell:1999ew}\footnote{
This leads to instabilities
which may be ameliorated for low energy effective theories
 with a sufficiently low
cutoff \cite{Carroll:2003st}.}.

In this note we consider restrictions on the dark energy equation of state
suggested by the generalized second law of thermodynamics (GSL)
\cite{Bekenstein:1973ur,Bekenstein:1974ax,Gibbons:1977mu,Gibbons:1976ue,
Unruh:1982ic,Davies:1987ti,Davies:1988dk},
\begin{equation}
\label{gsl}
dS=dS_{\rm{mat}}+2\pi dA_{\textrm{H}}\geq 0\ ,
\end{equation}
where $S_{\rm{mat}}$ is the entropy of matter and $A_{\rm{H}}$
is the horizon area (we use units with
$\hbar=c=8\pi G=1$). For vanishing $dA_{\textrm{H}}$,
Eq. \eqref{gsl} is the well known second law of ordinary thermodynamics.

How can one understand the need for the additional contribution from horizons?
This has first been discussed by Bekenstein for the case of a black hole
\cite{Bekenstein:1973ur}.
If matter carrying entropy falls into a black hole it is hidden from the
outside observer by its horizon, the surface that separates the
regions from which light rays can/cannot reach an infinitely distant observer. 
Classically, the black hole appears as a very well ordered object, fully
characterized by its mass, charge and angular momentum.
Naively, this should correspond to a state of very low entropy.
The outside observer would therefore conclude that the observable 
entropy decreases, $dS_{\rm{mat}} < 0$, seemingly in violation of the 
second law of thermodynamics. It is very intriguing, however, that
the horizon of a black hole grows when matter is added. Hence, one may
conjecture that a generalized second law, Eq. \eqref{gsl}, holds. Indeed,
the ignorance of the observer, and therefore the entropy he affiliates with 
the system, increases because he cannot see what is behind the horizon.

Horizons, similar to the one of a black hole, also appear in cosmology in 
case of accelerated expansion, most notably for de Sitter space \cite{suss}.
Shortly after the work of Bekenstein, Gibbons and Hawking suggested 
that also the future horizon of de Sitter space contributes to the total entropy
in the same way as the black hole event horizon \cite{Gibbons:1977mu},
which naturally leads to a generalized second law, Eq. \eqref{gsl},
also for de Sitter space \cite{Davies:1987ti,Davies:1988dk}.
Now the observer is inside the horizon and the horizon entropy represents
the lack of information about the outside region which he cannot see.
This conjecture has been examined in a variety of cases
\cite{Davies:1987ti,Davies:2003me,Davis:2003ye, buchjj}. However, the
cosmological situations are more delicate than the case of a black hole.
Except for de Sitter space, the horizon is not stationary, which implies
a departure from thermal equilibrium. This is related to an apparent
ambiguity in the choice of the ``horizon'' for which the GSL may hold.
Consider the following three possibilities, which all have identical area for
de Sitter space (we always assume spherical symmetry):
\begin{itemize}
\item{}The future or event horizon, which separates the regions 
from where light rays can/cannot reach an observer located at the center.
\item{}The Hubble or apparent horizon, which is the surface moving 
away from a centrally located observer with the speed of light.
\item{}The boundary of the causal region, with which an observer can 
communicate by sending a light ray and receiving the returned 
signal\footnote{Here, we measure the area of the surface at which the last light rays that asymptotically return to the oberver are reflected. This surface is the future horizon at a later time. Since in de Sitter space the area of the future horizon is constant the areas of the different horizons are equal.}.
\end{itemize}
All of these surfaces have appealing and less appealing features. The future
horizon, for example,  is a true horizon which separates regions from which a central observer
can/cannot receive information. However, it is not local in time and 
requires knowledge of the complete future evolution of the universe. In 
particular, one needs to know the equation of state until the infinite
future to calculate the future horizon. The Hubble horizon, on the other 
hand, depends only on the current state of the universe. Yet, it is not a true 
horizon; in many situations one can receive light rays from outside the
current Hubble horizon.

A related ambiguity concerns the volume used to calculate the matter entropy 
appearing in Eq. \eqref{gsl}. Possible choices include (cf. \cite{suss,bousso})
\begin{itemize}
\item{}The space-like volume ``inside'' the horizon.
\item{} The light-like hypersurface defined by light rays starting from the horizon 
\mbox{going ``in''}\footnote{If one wants to ensure that the light rays reach the centrally located observer one may have to start a tiny bit away from the horizon.}.
\item{} The light-like hypersurface given by light rays originating in the 
center.
\end{itemize}
Fortunately, the results for the different types of volumes do not differ, 
as long as we assume that no matter entropy is generated, as we will in 
this note.

From the above discussion it is clear that for general cosmological situations 
the status of the GSL is that of a conjecture backed up by some examples, 
where essential aspects remain to be clarified. Nevertheless, in the following
we will go ahead, apply the GSL and see what it can tell us about
the dark energy equation of state. To be explicit we will discuss two 
versions of the GSL, with the future horizon and the Hubble horizon,
respectively (for more details see \cite{buchjj}).
Furthermore, we will restrict the discussion to a flat universe as suggested by
observations.

As anticipated in \cite{Davies:1988dk,Brustein:1999ua,Izquierdo:2005ya}, 
and shown in some detail later on, superaccelerated expansion resulting from 
$w< -1$ typically is in conflict with Eq. \eqref{gsl}. The basic reason
is very simple. Consider the case where the entropy of the horizon is
the dominating contribution to the total entropy\footnote{We will see that 
this is a reasonable assumption for the late universe, but violating this 
assumption is also one possibility to get a phantom
cosmology in agreement with Eq. \eqref{gsl} (cf. \cite{buchjj}).}. 
During a phase of accelerated expansion the distance to the
horizon is $\sim \frac{1}{H}$ (future horizon and Hubble horizon are roughly 
of the same size) and its area is $A_{\textrm{H}}\sim\frac{1}{H^2}$. Since
the superaccelerated expansion is characterized by $\dot{H}>0$, it implies
$\dot{A}<0$. With $dS_{\textrm{mat}}\approx 0$ this is in contradiction
with \eqref{gsl}.
In the following we will illustrate this point and render it more precise.

We start in the next section by giving
an example which demonstrates the validity of the GSL for ``ordinary'' matter.
In section \ref{phantom} we then study models with different equations of 
state. Finally, we summarize and conclude in section \ref{conclusions}.

\section{An example for the GSL at work} \label{gslatwork}
To gain some confidence in the generalized second law let us briefly review a
simple  example: a flat universe filled with radiation and a
cosmological constant. For the metric
\begin{equation}
ds^2 = dt^2 - a(t)^2 \left(dr^2 + r^2 d\Omega^2\right)\ ,
\end{equation}
the proper distance to the future horizon is given by
\begin{equation}
\label{distance}
D_{\textrm{FH}}(t)=a(t) \int^{\infty}_{t}\frac{dt'}{a(t')}\ ,
\end{equation}
which leads to the horizon entropy (area)
\begin{equation}
\label{horizon-entropy}
S_{\textrm{FH}}=2\pi A_{\textrm{FH}}(t)=8\pi^2 D^{2}_{\textrm{FH}}(t)\ .
\end{equation}
The Hubble horizon is simply given by
\begin{equation}
D_{\rm{Hub}}(t)=\frac{1}{H(t)}\ ,
\end{equation}
and the corresponding entropy would be
\begin{equation}
\label{horizon-entropy2}
S_{\textrm{Hub}}=2\pi A_{\textrm{Hub}}(t)=8\pi^2 \frac{1}{H^{2}(t)}\ .
\end{equation}

The evolution of the scale factor $a(t)$ is determined by the Friedmann 
equations
\begin{equation}
\label{friedmann}
3H^2=\rho\ , \quad \dot{\rho}+3H(\rho+p)=0\ .
\end{equation}
In our simple example the energy density is the sum 
$\rho=\rho_{\textrm{R}}+\rho_{\Lambda}$. Here, radiation and cosmological 
constant correspond to perfect fluids with the equations of state
\begin{equation}
p_{\textrm{R}}=w_{\textrm{R}}\rho_{\textrm{R}}=\frac{1}{3}\rho_{\textrm{R}}\ ,
\quad\rm{and}\quad
p_{\Lambda}=w_{\Lambda}\rho_{\Lambda}=-\rho_{\Lambda}=-\Lambda\ ,
\end{equation}
for the radiation and the cosmological constant component, respectively. 
For radiation (photons), energy and entropy density are determined by the
temperature,
\begin{equation}
\rho_{\textrm{R}}=\sigma T^4\ , \quad
s_{\textrm{R}}=\frac{4}{3}\sigma T^3\ , \quad
\sigma=\frac{\pi^2}{15}\ .
\end{equation}
The total entropy inside the horizon\footnote{We take a space-like volume 
and the equal-time is specified by an
observer resting with respect to the fluid.}
is then given by
\begin{equation}
\label{radiation-entropy}
S_{\textrm{R}}=\frac{4\pi}{3} D^3_{\textrm{H}}(t) s_{\textrm{R}}(t)
=\frac{16\pi}{9}\sigma T^3_{0}\frac{D^3_{\textrm{H}}(t)}{a^3(t)}\ .
\end{equation}
For the last equality we have used that for a gas of massless particles 
$T=\frac{T_{0}}{a(t)}$, with $T_{0}=T(t_{0})$ and $a(t_{0})\equiv 1$.

\begin{figure*}[t]
\begin{center}
\scalebox{0.8}[0.8]{
\begin{picture}(190,140)(40,0)
\includegraphics[width=9.5cm]{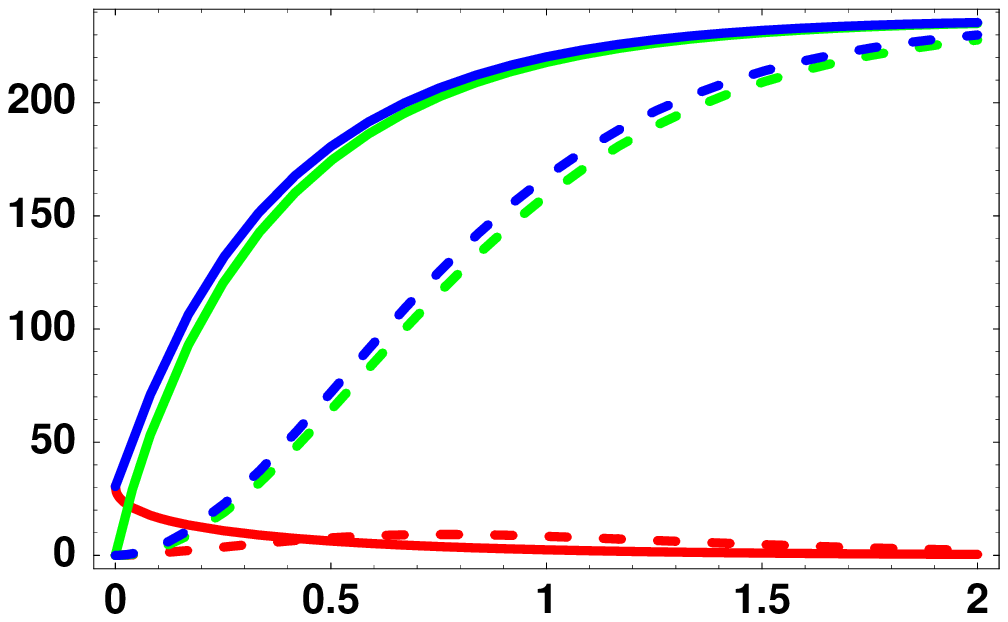}
\Text(-40,-20)[c]{\scalebox{1.6}[1.6]{$t$}}
\Text(-280,150)[c]{\scalebox{1.6}[1.6]{$S$}}
\end{picture}
}
\end{center}
\caption{Time evolution of the entropy for a universe filled with radiation 
and a cosmological constant ($\Lambda=1$). Matter entropy inside the horizon 
(red) and horizon entropy (green) add up to the total entropy (blue). 
Solid lines are for the future horizon, dashed lines for the Hubble horizon.} 
\label{radexample}
\end{figure*}

As shown in \cite{Davies:2003me}, the Friedmann equations can be solved
analytically in this case, yielding
\begin{equation}
a(t)=\left(\frac{\sigma T^{4}_{0}}{\Lambda}\right)^{\frac{1}{4}}
\left(\sinh(2\sqrt{\frac{\Lambda}{3}}\,t)\right)^{\frac{1}{2}}.
\end{equation}
Inserting this result for the scale factor into Eqs. \eqref{horizon-entropy}, 
\eqref{radiation-entropy} we obtain
\begin{eqnarray}
\label{eventexample}
&&\!\!\!\! S_{\textrm{FH}}=\frac{6\pi^2}{\Lambda}\sinh(x) J^{2}(x)\ ,
\\\nonumber
&&\!\!\!\! S^{\rm{FH}}_{\textrm{R}}=\frac{2\pi}{\sqrt{3}}
\sigma^{\frac{1}{4}}\Lambda^{-\frac{3}{4}}J^{3}(x)\ ,
\\\nonumber
&&\!\!\!\! x=2\sqrt{\frac{\Lambda}{3}}t, \quad 
J(x)=\int^{\infty}_{x}\frac{dx}{(\sinh(x))^{\frac{1}{2}}}\ ,
\end{eqnarray}
if we take the future horizon, and
\begin{eqnarray}
\label{hubbleexample}
&&\!\!\!\! S_{\textrm{Hub}}=\frac{24\pi^{2}}{\Lambda}(\tanh(x))^{2}\ ,
\\\nonumber
&&\!\!\!\! S^{\rm{Hub}}_{\textrm{R}}=\frac{16}{\sqrt{3}}\pi
\sigma^{\frac{1}{4}}\Lambda^{-\frac{3}{4}}
(\tanh(x))^{\frac{3}{2}}(\cosh(x))^{-\frac{3}{2}}
\end{eqnarray}
for the Hubble horizon.
We note that the entropies are independent of the initial temperature $T_{0}$ due to our choice of the initial time $t_{0}$.
The results are plotted in Fig.
\ref{radexample}.

\begin{figure*}[t]
\begin{center}
\subfigure[]{\scalebox{0.75}[0.75]{
\begin{picture}(190,140)(40,0)
\includegraphics[width=9.5cm]{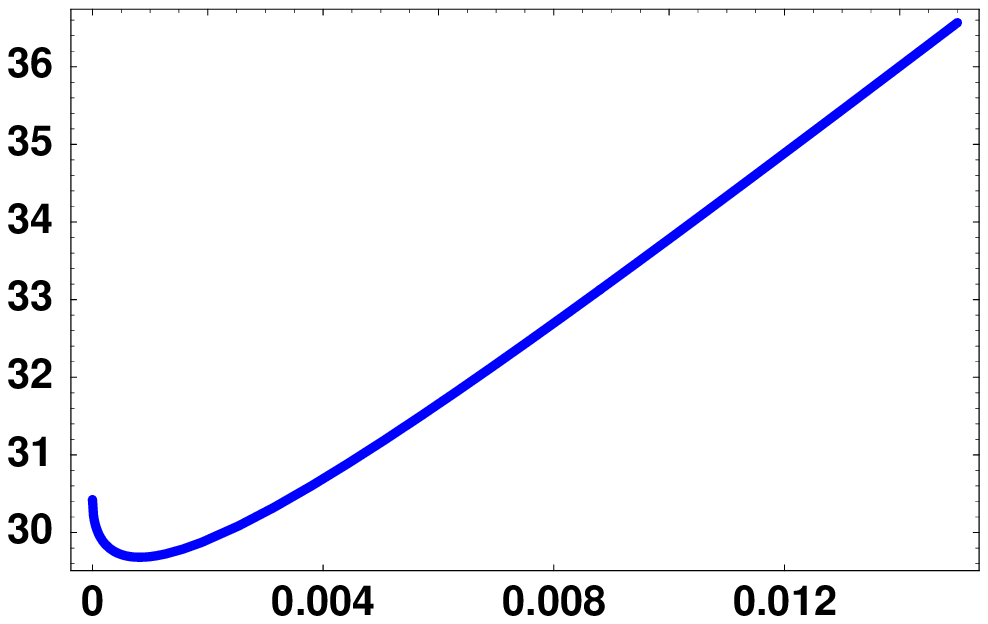}
\Text(-40,-20)[c]{\scalebox{1.6}[1.6]{$t$}}
\Text(-280,150)[c]{\scalebox{1.6}[1.6]{$S$}}
\end{picture}
}\label{highenergydrop}
}
\hspace{2.5cm}
\subfigure[]{\scalebox{0.75}[0.75]{
\begin{picture}(190,160)(40,0)
\includegraphics[width=9.5cm]{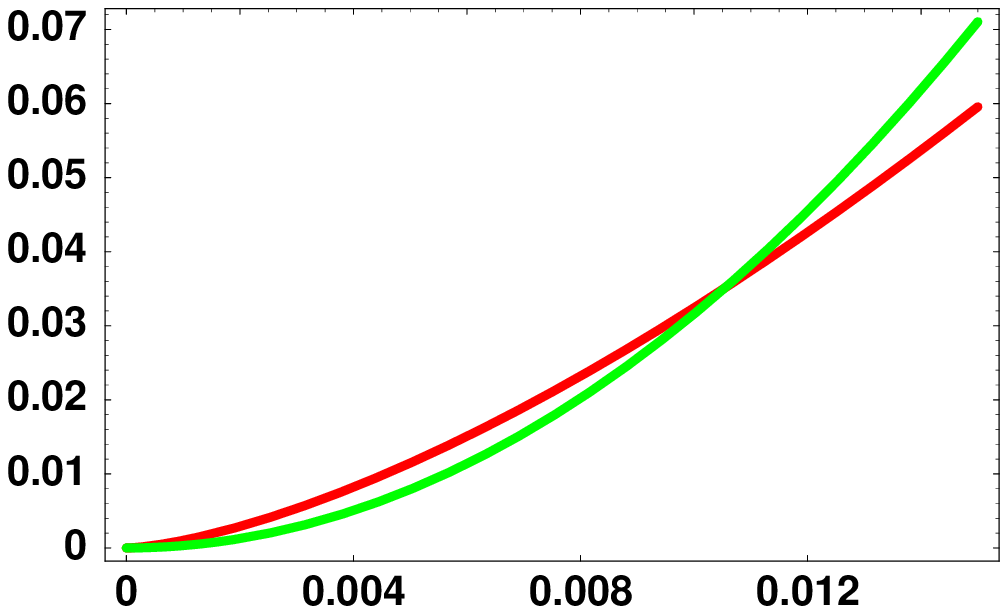}
\Text(-40,-20)[c]{\scalebox{1.6}[1.6]{$t$}}
\Text(-280,150)[c]{\scalebox{1.6}[1.6]{$T$}}
\end{picture}
}\label{entropybound}
}
\end{center}
\vspace{-0.8cm} \caption{Left panel, total entropy for very early times 
(parameters as in Fig. \ref{radexample})
using the future horizon.
In the right panel the total matter entropy inside a Hubble volume (red) is 
compared to the maximal entropy allowed by the Bousso bound \cite{bousso} 
(green).} \label{highenergydroptotal}
\end{figure*}

\begin{figure*}[t]
\begin{center}
\subfigure[]{\scalebox{0.75}[0.75]{
\begin{picture}(190,140)(40,0)
\includegraphics[width=9.5cm]{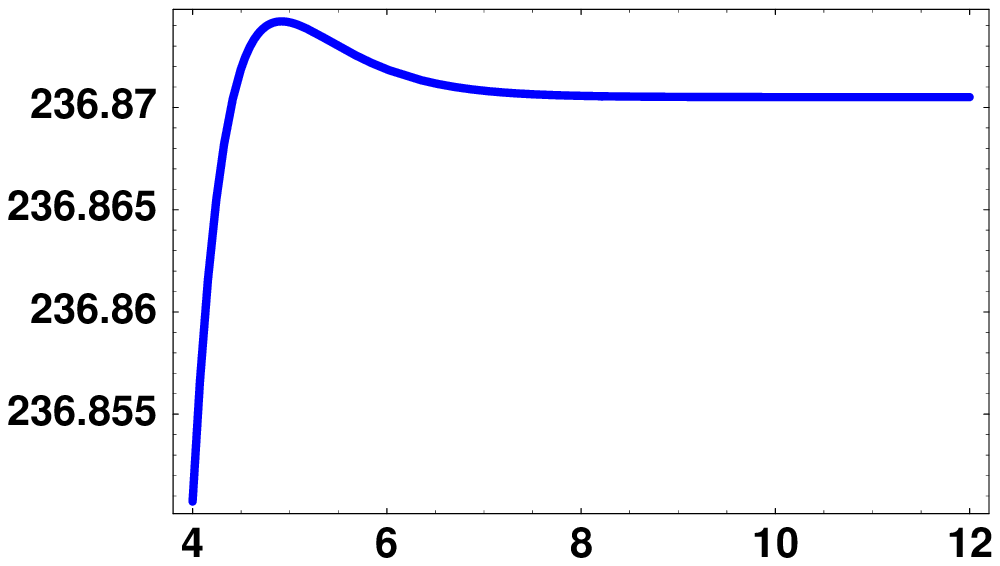}
\Text(-40,-20)[c]{\scalebox{1.6}[1.6]{$t$}}
\Text(-280,150)[c]{\scalebox{1.6}[1.6]{$S$}}
\end{picture}
}\label{lowenergydrop}
}
\hspace{2.5cm}
\subfigure[]{\scalebox{0.75}[0.75]{
\begin{picture}(190,140)(40,0)
\includegraphics[width=9.5cm]{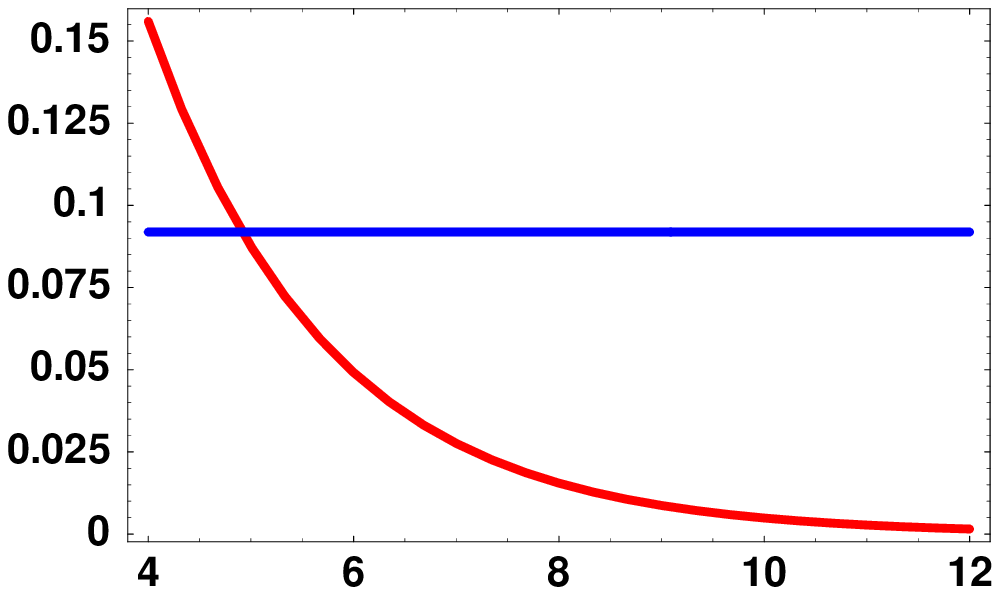}
\Text(-40,-20)[c]{\scalebox{1.6}[1.6]{$t$}}
\Text(-280,150)[c]{\scalebox{1.6}[1.6]{$T$}}
\end{picture}
}\label{horizontemperature}
}
\end{center}
\vspace{-0.8cm} \caption{Left panel: evolution of the total entropy 
for future horizon at late times; for the Hubble horizon the picture is 
very similar (parameters as in Fig. \ref{radexample}). Right panel: comparison
of the radiation temperature $T=T_{0}/a(t)$ (red) with the horizon temperature
$1/(2\pi D_{\rm{H}}(t))$ (blue).} \label{lowenergydroptotal}
\end{figure*}

Independent of our choice for the horizon, i.e. for future horizon as 
well as Hubble horizon, 
the GSL, Eq. \eqref{gsl}, appears to work. The blue curves, which represent 
the sum of matter entropy inside the horizon and horizon entropy, both seem 
to increase with time monotonously. However, with the help of a 
``magnifying glass'' one can spot a small decrease in the total entropy at 
very early and very late times for the future horizon,
Fig. \ref{highenergydrop} and Fig. \ref{lowenergydrop}. In case of the 
Hubble horizon there is no decrease at early times.

Consider first the small drop in the total entropy at early times, 
Fig. \ref{highenergydrop}. We don't have to worry about this decrease for
two reasons. First, one can easily check that at these early times 
the temperature is well above the Planck scale, so that we cannot expect the
formula Eq. \eqref{radiation-entropy} for the matter entropy to hold.
Second, the covariant entropy bound \cite{bousso} is not fulfilled at these
early times. This is depicted in Fig. \ref{entropybound} where the total
matter entropy inside a Hubble volume is compared with the maximally
allowed entropy inside this volume \cite{bousso}.
In the region where the entropy decreases the covariant entropy bound is 
clearly violated, and the used expression for the entropy is therefore
inapplicable.

Let us now turn to the drop in entropy at very late times, 
Fig. \ref{lowenergydrop}\footnote{Note that the time between Planck scale 
and Hubble scale temperatures, where the GSL is valid, is rather small 
due to the large value chosen for the cosmological constant, 
$\Lambda=1$ in Planck units.}.
At the corresponding temperatures the 
typical wavelength of photons contributing to the matter entropy,
$\lambda\sim1/T$, is bigger that the horizon size at this time. Since
the wavelength doesn't fit into the horizon anymore, the flat space 
relation Eq. \eqref{radiation-entropy} is no longer applicable. 
In Fig. \ref{horizontemperature} the radiation temperature 
$T=T_{0}/a(t)$ is compared to the ``horizon temperature'' 
$1/(2\pi D_{\rm{H}}(t))$, which can roughly be thought of as the smallest 
possible temperature in an expanding universe. Again, the drop
in entropy occurs when Eq. \eqref{radiation-entropy} is no longer applicable.

In summary, we find that the GSL works independent of our choice of horizon 
(future horizon or Hubble horizon). Indeed, apparent violations of the GSL
can be easily understood as consequence of the inapplicability of the 
flat space expression Eq. \eqref{radiation-entropy} for the radiation
entropy.

\section{Decreasing horizon entropy in phantom models}\label{phantom}
Having gained some confidence in the GSL by studying the example in the 
previous section, let us now move on to a more interesting situation.
Consider the following simple model for the equation of state,
\begin{equation}
\label{eqstate}
w(\rho_{\rm{DE}})=-1+\Delta w \,\Theta(\rho_{\textrm{max}}-\rho_{\rm{DE}}).
\end{equation}
For $\Delta w <0$, the dark energy has phantom behaviour at early times.
From the Friedmann equations one finds that
the energy density $\rho_{\rm{DE}}$ grows until it reaches the maximal
density $\rho_{\textrm{max}}$ which acts like a cosmological constant.
In this way the ``big rip'' singularity is avoided which occurs for a constant 
equation of state $w(\rho_{\rm{DE}})<-1$. 
$\Delta w=0$ is the case of a cosmological constant. For
$\Delta w >0$ one obtains a quintessence-like model with constant equation of 
state at late times. At early times the maximal density 
$\rho_{\textrm{max}}$ is not exceeded, avoiding the initial singularity.
For $\Delta w < {2\over 3}$ one has an accelerated expansion of the universe
which is associated with the formation of a future horizon.

\begin{figure*}[t!]
\begin{center}
\subfigure[]{\scalebox{0.76}[0.76]{
\begin{picture}(190,140)(40,0)
\includegraphics[width=9.5cm]{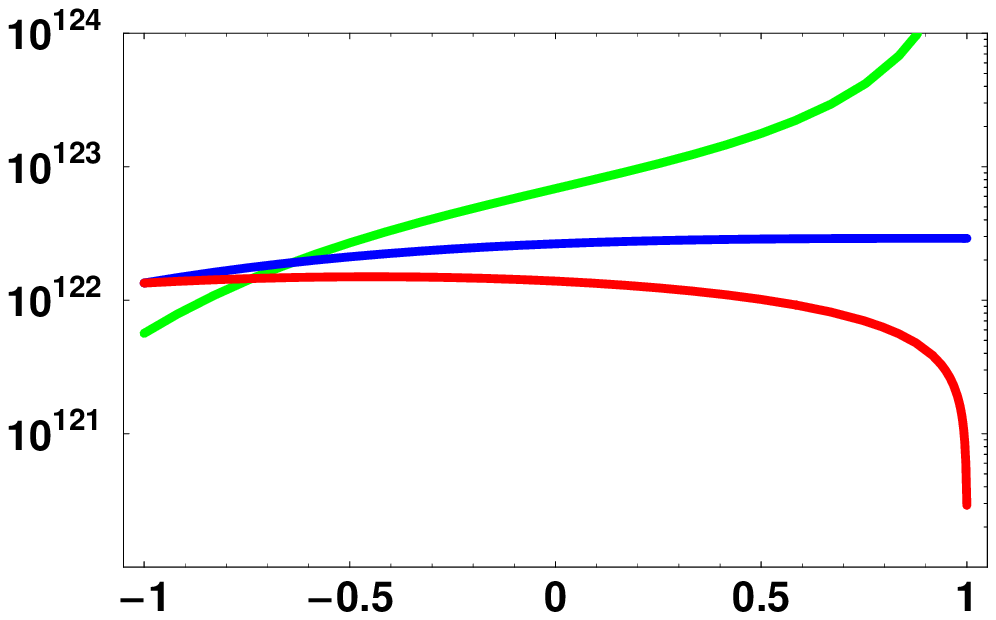}
\Text(-40,-20)[c]{\scalebox{1.6}[1.6]{$\eta/\eta_{\textrm{end}}$}}
\Text(-280,150)[c]{\scalebox{1.6}[1.6]{$S$}}
\end{picture}
}\label{vergleich}
}
\hspace{2.5cm}
\subfigure[]{\scalebox{0.75}[0.75]{
\begin{picture}(190,140)(40,0)
\includegraphics[width=9.5cm]{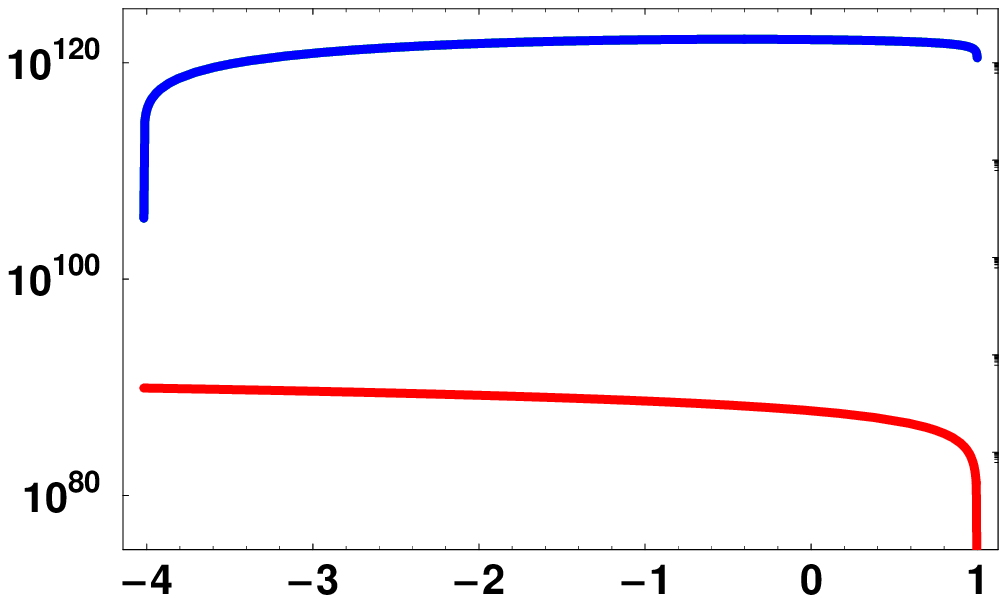}
\Text(-40,-20)[c]{\scalebox{1.6}[1.6]{$\eta/\eta_{\textrm{end}}$}}
\Text(-280,150)[c]{\scalebox{1.6}[1.6]{$S$}}
\end{picture}
}\label{radiation}
}
\end{center}
\vspace{-0.8cm} \caption{Left panel: evolution of the total  entropy
$S=S_{\textrm{rad}}+S_\textrm{H}$ with (conformal) time $\eta$ (rescaled 
by $\eta_{\rm{end}}$, the conformal time at 
$t=\infty$). Parameters:
$\Omega_{\textrm{mat}}=0.3$, $\Omega_{\textrm{DE}}=0.7$, 
$T_\textrm{rad}=2.7K$. The green curve represents a quintessence universe 
($\Delta w=0.25$), the blue one a cosmological constant ($\Delta w=0$)
and the red one a phantom universe ($\Delta w=-0.25$); 
$\rho_{\rm{max}}=100\rho_{\rm{today}}$. Right panel:
comparison of horizon entropy (blue) and radiation entropy (red)
for the phantom model. 
The radiation entropy is negligible at late times. In all cases the future
horizon has been chosen; the results for the Hubble horizon are 
qualitatively similar.} \label{complete}
\end{figure*}

In Fig.~\ref{complete} the evolution of the total entropy 
$S=S_{\textrm{rad}}+S_\textrm{H}$ is shown for a quintessence model, a 
cosmological constant and a phantom model, respectively. In addition to the 
dark energy, described by the equation of state given in Eq.~\eqref{eqstate}, 
we have added a component of dark matter ($w=0$, $S_{\textrm{DM}}=0$) and a
radiation component ($w=\frac{1}{3}$). $\eta$ is the conformal time 
($d\eta = dt/a(t)$), with $\eta(t_0) = 0$. 
The energy densities are fixed by
their current values, $T_\textrm{rad}=2.7K$, 
$H_{0}=70\ \rm{km}/(\rm{s}\,\rm{Mpc})\sim 6\times 10^{-61} \,M_{\rm{P}}$ 
and $\Omega_{\textrm{DE}}/\Omega_{\textrm{DM}}=0.7/0.3$, where 
$\Omega_{DE}$ and $\Omega_{\rm{DM}}$ are the fractions of the total energy 
density contributed by dark energy and dark matter, respectively.
The total entropies of the three models are plotted in 
Fig.~\ref{vergleich}. The entropy increases both for quintessence and
for the cosmological constant. In the quintessence model the entropy increases
without bound, as the horizon continues to grow with time. In both models the 
GSL is fulfilled. On the contrary, for the phantom model the total entropy 
first increases but then decreases in violation of the GSL. In particular
today, at $\eta = 0$, the phantom model is inconsistent with the GSL.

In Fig.~\ref{radiation} the horizon entropy for the phantom model is
compared with the radiation entropy inside the horizon. Most of the time
the horizon entropy is larger than the matter entropy by many orders of 
magnitude. One easily verifies that this holds for all three models. 
Moreover, the matter entropy inside the horizon decreases with time. 
In fact, one can infer from the Friedmann equations that accelerated expansion
leads to a decrease in entropy inside the horizon for all components 
which expand adiabatically such as radiation. Hence, as long as we consider 
only adiabatic expansion where no matter entropy is generated for any
component, the GSL is always violated as soon as the horizon and therefore 
the horizon entropy begins to shrink.

What can we learn from this? 
If $\rho_{\rm{max}}$ is sufficiently large, $\rho_{\rm{max}}>\rho_{0}$, 
where $\rho_{0}$ is the current total energy density of the universe, 
then the GSL is violated for all $\Delta w<0$.
The more conservative requirement $\rho_{\rm{max}}=\rho^{\rm{DE}}_{0}$,
where $\rho^{\rm{DE}}_{0}$ is the current dark energy density, would
allow $\Delta w<0$ only in the past, with a cosmological constant in the
future making the current era very special. 
In our simple model one then finds that consistency with the GSL requires
$\Delta w >-1.8$ if we use the future horizon and $\Delta w >-0.43$ for 
the Hubble horizon.

\section{Summary and Conclusions} \label{conclusions}
Generically, gravity leads to the formation of horizons which separate
causally disconnected regions. The second law of thermodynamics, which
applies to closed systems, then has to be modified. The conjecture
of a generalized second law (GSL), which applies to the sum of ordinary
matter entropy and horizon entropy, is well established for black holes
which represent stationary systems. 
For non-stationary systems, as they appear in cosmology, the status of the 
GSL is much more speculative and several questions, like the proper choice
of the horizon, remain to be clarified. Yet, as demonstrated by our simple 
example of a universe made of radiation and a cosmological constant, the GSL 
appears to work in cosmological situations, too.

Keeping the above caveats in mind it is nevertheless interesting to apply 
the GSL and ask what it can tell us about the dark energy equation of state. 
In general, both quintessence, $w_{\rm{DE}}>-1$, and a cosmological constant,
$w_{\rm{DE}}=-1$, are consistent with a GSL. On the contrary, long phases of 
a phantom  equation of state, $w_{\rm{DE}}<-1$, typically lead to a decrease 
of the total entropy. Short phases with a phantom equation of state might
be allowed.

In conclusion, further studies on the validity of a generalized second law 
of thermodynamics are an important theoretical challenge. Our simple example 
of the dark energy equation of state already illustrates its potential as a 
tool for cosmology.

\section*{Acknowledgements}
The authors would like to thank the organizers of the QUARKS 2006 for a 
wonderful conference in the stimulating environment of St. Petersburg.

\end{document}